\documentclass[conference]{IEEEtran}
\usepackage{algorithm}
\usepackage{array}
\usepackage[caption=false,font=normalsize,labelfont=sf,textfont=sf]{subfig}
\usepackage{textcomp}
\usepackage{stfloats}
\usepackage{url}
\usepackage{verbatim}
\usepackage{graphicx}
\usepackage{cite}
\usepackage{amssymb}

\usepackage{amsmath,amssymb,amsfonts}
\usepackage{algpseudocode}
\usepackage{multirow}
\usepackage{xcolor}
\usepackage{xfrac}
\usepackage{pbox}
\usepackage{amsmath}

\begin{document}

\newcommand{\pamEightPowerFunction}{PAM3\_Power}
\newcommand{\pamEightData}{twoLines} 
\newcommand{\flagIndex}{flagIndex}

\algnewcommand\algorithmicswitch{\textbf{switch}}
\algnewcommand\algorithmiccase{\textbf{case}}
\algnewcommand\algorithmicassert{\texttt{assert}}
\algnewcommand\Assert[1]{\State \algorithmicassert(#1)}%

\algdef{SE}[SWITCH]{Switch}{EndSwitch}[1]{\algorithmicswitch\ #1\ \algorithmicdo}{\algorithmicend\ \algorithmicswitch}%
\algdef{SE}[CASE]{Case}{EndCase}[1]{\algorithmiccase\ #1}{\algorithmicend\ \algorithmiccase}%
\algtext*{EndSwitch}%
\algtext*{EndCase}%

\newdimen{\algindent}
\setlength\algindent{1.2em}          
\algnewcommand\LeftComment[2]{%
\hspace{#1\algindent}$\triangleright$ {#2} \hfill %
}
\IEEEoverridecommandlockouts
\IEEEaftertitletext{\vspace{-2.75\baselineskip}}


\title{%
		\vspace{-2em}
		\begin{center}
				\small This is the authors' copy of the paper to appear in Proceedings of the 20th International Conference on Synthesis, Modeling, Analysis and Simulation Methods, and Applications to Circuit Design (SMACD 2024)
			\end{center}
		\vspace{0.3em}
		Low-Power Encoding for PAM-3 DRAM Bus
}

\author{
\IEEEauthorblockN{Jonghyeon Nam}
\IEEEauthorblockA{
\textit{Pohang University of Science and Technology} \\
Pohang, Republic of Korea\\
jhnam99@postech.ac.kr}
\and
\IEEEauthorblockN{Jaeduk Han}
\IEEEauthorblockA{
\textit{Hanyang University}\\
Seoul, Republic of Korea\\
jdhan@hanyang.ac.kr}
\and
\IEEEauthorblockN{Hokeun Kim}
\IEEEauthorblockA{
\textit{Arizona State University} \\
Tempe, AZ, United States\\
hokeun@asu.edu}
}



\maketitle

\begin{abstract}
The 3-level pulse amplitude modulation (PAM-3) signaling is expected to be widely used in memory interfaces for its greater voltage margins compared to PAM-4.
To maximize the benefit of PAM-3, we propose three low-power data encoding algorithms: \emph{PAM3-DBI}, \emph{PAM3-MF}, and \emph{PAM3-SORT}.
With the DRAM memory traces from the gem5 computer architecture simulator running benchmarks, we evaluate the energy efficiency of our three PAM-3 encoding techniques.
The experimental results show the proposed algorithms can reduce termination power for high-speed memory links significantly by 41\% to 90\% for benchmark programs.
\end{abstract}


\section{Introduction}
Due to the latest developments in high-performance computing systems for data-centric applications, there has been a significant rise in the processing and communication demands of processors and memory systems.
The per-pin data rates for the computing and graphic memory interfaces are anticipated to consistently increase~\cite{isscc_trends}.
However, as the speed of the DRAM interface increases, signal losses and inter-symbol interference (ISI)~\cite{mem_interface} occur due to non-ideal factors such as skin effect and dielectric loss, limiting the growth of data rate in line with system requirements. 

High-speed analog equalizers can mitigate the effects of ISI but with the cost of significantly increased power consumption at high frequencies. 
Therefore, recent advancements have introduced modulation techniques such as 3 or 4-level pulse-amplitude modulations (PAM-3 or PAM-4) to enhance the data rate without increasing the operating frequency~\cite{GDDR6X, PAM3}.

Especially, the PAM-3 scheme has been extensively explored in recent studies on memory interfaces~\cite{PAM3, PAM3_2, PAM3_3} as it offers greater voltage margins than PAM-4 signaling.
To achieve the maximum benefit of the PAM-3 scheme, this paper proposes novel algorithms for PAM-3 encoding, significantly advancing the previous work~\cite{su2022energy} that proposed and analyzed data-encoding methods for PAM-4.

\section{PAM-3 Encoding Algorithms}

\begin{figure}
\centering
\includegraphics[width=0.75\columnwidth]{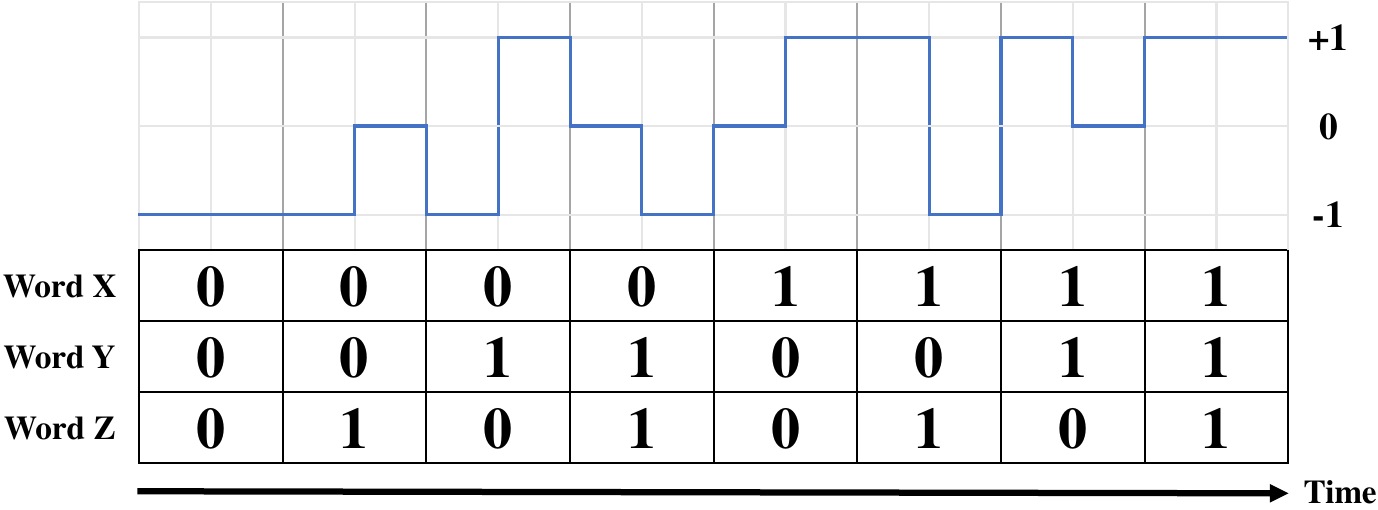}
\caption{PAM-3 signal modulation method. Each column of binary symbols 000, 001, 010, ..., 111 from three 8-bit word lines (X, Y, and Z) are mapped to PAM-3 signals [-1, -1], [-1, 0], [-1, +1], ..., [+1, +1] of two PAM-3 lines.}
\label{fig:pam3_signal_modulation}
\end{figure}

We propose three energy-efficient encoding algorithms for PAM-3, namely, (A) \emph{PAM3-DBI}, (B) \emph{PAM3-MF}, and (C) \emph{PAM3-SORT}. 
\emph{DBI} stands for ``\emph{Data Bus Inversion},'' 
\emph{MF} stands for ``\emph{Most Frequent},'' and
\emph{SORT} means we ``\emph{sort} signals by their frequency.''
For binary input to PAM-3 conversion, we use the mapping presented in \figurename~\ref{fig:pam3_signal_modulation}.
For power consumption information of each PAM-3 signal, we use the numbers in~\tablename~\ref{tab:pam3_table}.
Each algorithm has a specific method to convert the PAM-3 signals to reduce overall termination power.

\begin{table}
\centering
\caption{\label{tab: pam3-table}PAM-3 Power Consumption}
\large
\scalebox{0.7}{
\begin{tabular}{|c|ccc|}
\hline & \multicolumn{3}{c|}{PAM-3} 
\\ 
\hline Signal & \multicolumn{1}{c|}{-1} & \multicolumn{1}{c|}{0} & +1
\\
\hline \begin{tabular}[c]{@{}c@{}}Power \\ Consumption\end{tabular} & \multicolumn{1}{c|}{\Large $\frac{V^2_{DD}}{100}$}    & \multicolumn{1}{c|}{\Large $\frac{V^2_{DD}}{200}$} & 0   
\\ \hline
\end{tabular}}
\label{tab:pam3_table}
\end{table}

\begin{figure}
\centering
\includegraphics[width=0.62\columnwidth]{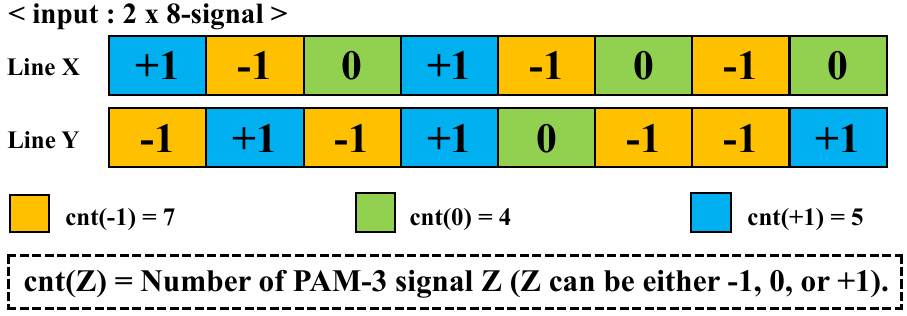}
\caption{An example for the definition of notation cnt(Z).}
\label{fig:cnt_z}
\end{figure}

\subsection{PAM3-DBI Encoding}

\begin{algorithm}
\footnotesize
  \centering
  \caption{PAM3-DBI Encoding}
  \label{alg:PAM3-DBI}
  \begin{algorithmic}[1]
  \State \LeftComment{0}{\Call{invert}{$data$}: Returns the given data bits with each bit inverted.}
  \State \LeftComment{0}{\Call{count}{$data$}: Returns $cnt(-1), cnt(0)$, and $cnt(+1)$, the numbers of PAM-3 signals, -1, 0, and +1 in the given data, respectively.}
  \Procedure{PAM3\_DBI\_Encoding}{$\pamEightData$}
    \State $invFlag \gets 0$
    \State $cnt(-1), cnt(0), cnt(+1) \gets \Call{count}{\pamEightData}$
    \If{$cnt(-1) > cnt(+1)$} \label{alg_line:pm3_dbi_cond}
            \State $\pamEightData \gets \Call{invert}{\pamEightData}$
            \State $invFlag \gets 1$
    \EndIf
    \State\Return $\pamEightData$, $invFlag$
  \EndProcedure
  \end{algorithmic}
\end{algorithm}

\IEEEpubidadjcol

Algorithm \ref{alg:PAM3-DBI} illustrates the PAM3-DBI encoding method. The PAM3-DBI encoding inverts each signal if the inverted signals reduce overall power consumption.
For brevity, we use the $cnt(Z)$ notation depicted in \figurename~\ref{fig:cnt_z}. $cnt(Z)$ represents the total number of each value in a signal composed of two lines.
Specifically, based on the power consumption parameters presented in \tablename~\ref{tab:pam3_table} and the values of $cnt(-1)$, $cnt(0)$, and $cnt(+1)$, we express the condition for converting each signal to reduce the overall power consumption as shown below:

\begin{equation}
\footnotesize
\frac{cnt(-1)}{100} + \frac{cnt(0)}{200} + cnt(+1) \times 0
\\
>   \frac{cnt(+1)}{100} + \frac{cnt(0)}{200} + cnt(-1) \times 0
\label{eq:PAM3-DBI_1}
\end{equation}

The left-hand side of (\ref{eq:PAM3-DBI_1}) is the power consumption of the given two signal lines, and the right-hand side of (\ref{eq:PAM3-DBI_1}) is the power consumption after applying signal-conversion on the given data.
(\ref{eq:PAM3-DBI_1}) can be simplified as follows: 

\begin{equation}
\footnotesize
\begin{aligned}
cnt(-1) > cnt(+1)
\end{aligned}
\label{eq:PAM3-DBI_2}
\end{equation}

This simplified inequality is used at line \ref{alg_line:pm3_dbi_cond} of Algorithm \ref{alg:PAM3-DBI}.

\subsection{PAM3-MF Encoding}
\label{sec:pam3-mf}

\begin{figure}
\centering
\includegraphics[width=0.75\columnwidth]{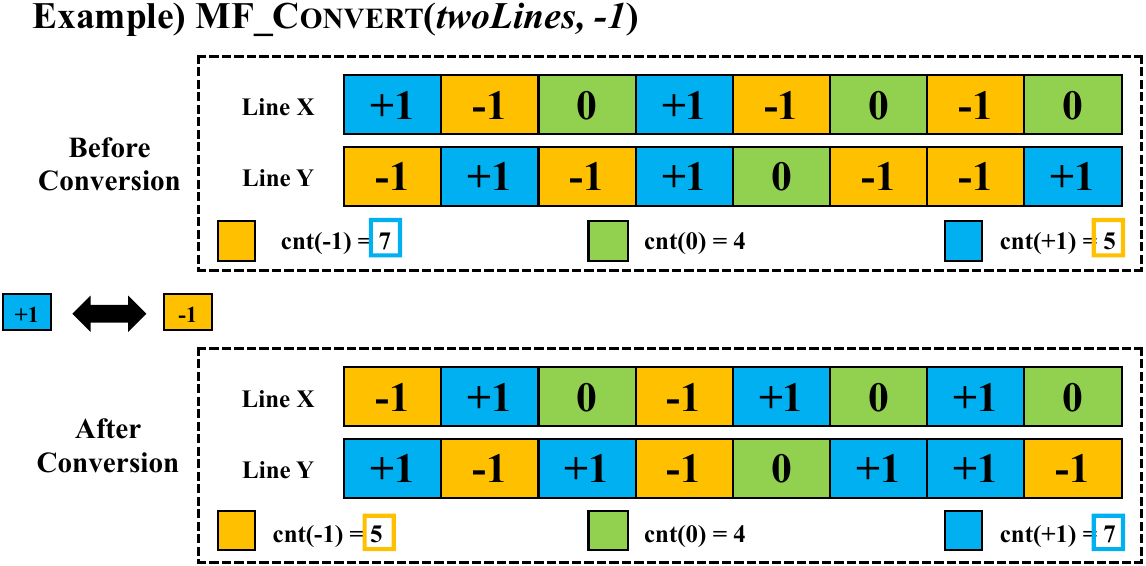}
\caption{An example operation of the helper procedure, \textsc{MF\_Convert}, used in PAM3-MF encoding (Algorithm \ref{alg:PAM3-MF}), where the given signal to be swapped with +1 signal is -1.}
\label{fig:mf_convert}
\end{figure}

\begin{algorithm}[tb]
  \footnotesize
  \centering
  \caption{PAM3-MF Encoding}
  \label{alg:PAM3-MF}
  \begin{algorithmic}[1]
  \State \LeftComment{0}{\Call{argmax}{$array$}: Returns the index of the max value in the given array.}
  \State \LeftComment{0}{\Call{MF\_Convert}{$data$, $signal$}: For the given data, swaps +1 symbols with the given signal.}\label{alg_line:mf_convert1}
  \Procedure{PAM3\_MF\_Encoding}{$\pamEightData$}
    \State $cnt(-1), cnt(0), cnt(+1) \gets \Call{count}{\pamEightData}$
    \State \LeftComment{0}{$mf\_signal$ is a 2-bit representation of the most frequent signal.}
    \State $mf\_signal \gets \Call{argmax}{[cnt(-1), cnt(0), cnt(+1)]}$
    \State $\pamEightData \gets \Call{MF\_Convert}{\pamEightData, mf\_signal}$\label{alg_line:mf_convert2}
    \State\Return $\pamEightData, mf\_signal$
  \EndProcedure
  \end{algorithmic}
\end{algorithm}

Algorithm \ref{alg:PAM3-MF} describes PAM3-MF, our second encoding algorithm proposed in this paper.
The PAM3-MF encoding algorithm is \emph{simple yet the most cost-effective} among the three proposed PAM-3 encoding algorithms. 

\figurename~\ref{fig:mf_convert} illustrates an example operation of \textsc{PAM3\_MF\_Convert}, defined at line \ref{alg_line:mf_convert1} of Algorithm \ref{alg:PAM3-MF} and called at line \ref{alg_line:mf_convert2}. In this example, two lines of PAM-3 signals and -1 are given as parameters of \textsc{PAM3\_MF\_Convert}.
The -1 signal is given as the signal to be swapped with the +1 signal as it is the most frequent among three symbols, i.e., $cnt(-1)$ is the largest among, $cnt(-1)$ $cnt(0)$, and $cnt(+1)$ (before conversion).
During conversion, -1 signals are replaced with +1 signals that have the lowest power consumption, resulting in two lines shown in ``after conversion'' in \figurename~\ref{fig:mf_convert}.


\begin{figure}
\centering
\includegraphics[width=0.77\columnwidth]{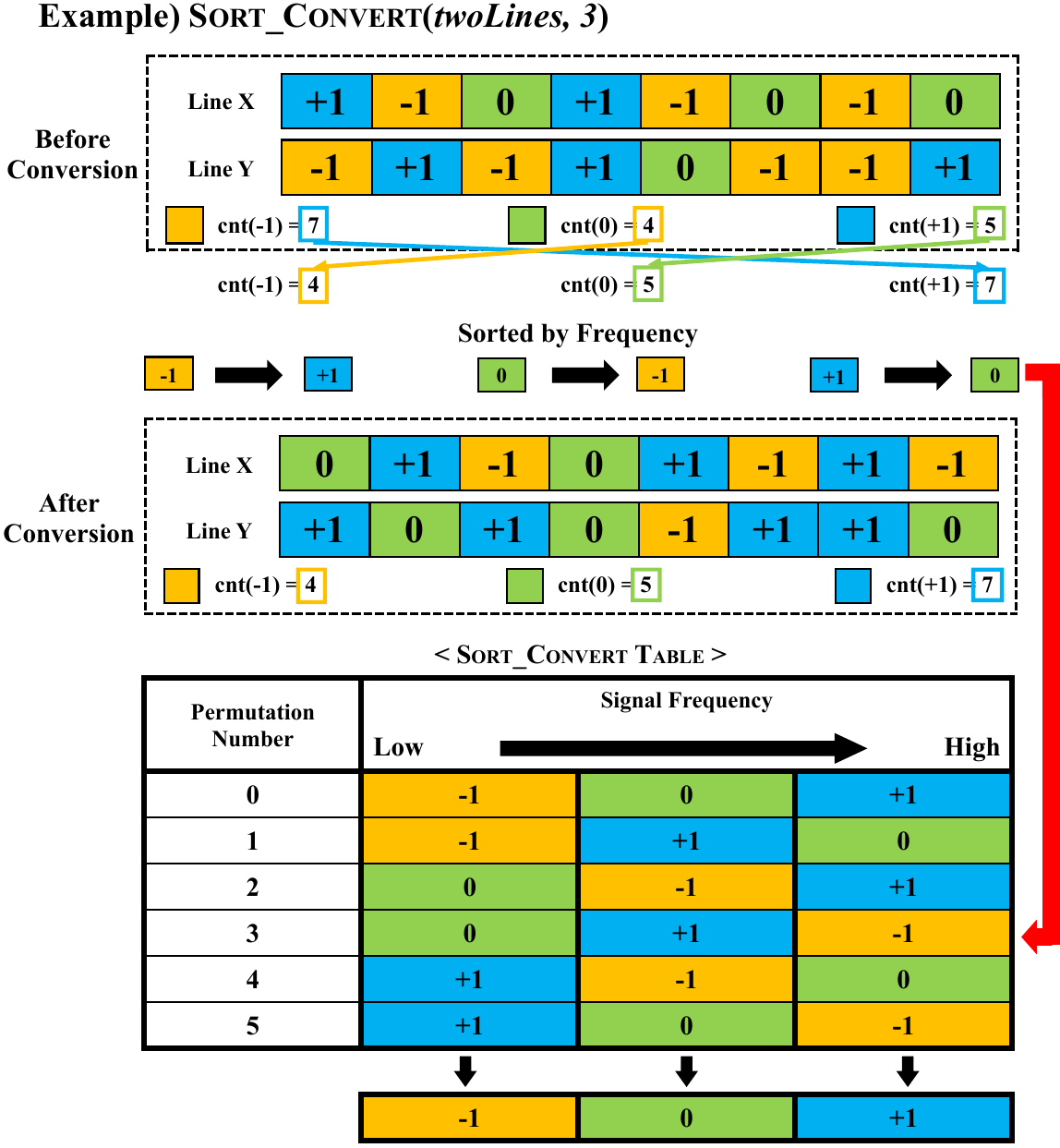}
\caption{An example process of sorting signals from the least frequent one to the most frequent one, assigning a permutation number for the sorted signals, and applying a helper procedure, \textsc{Sort\_Convert}, to convert the data for the given permutation number to achieve the minimum power consumption.
	(The permutation [0, 1, -1] is assigned the permutation number 3 in this example.)}
\label{fig:sort_convert}
\end{figure}

\subsection{PAM3-SORT Encoding}
\label{sec:pam3-sort}

\begin{algorithm}
  \footnotesize
  \centering
  \caption{PAM3-SORT Encoding}
  \label{alg:PAM3-SORT}
  \begin{algorithmic}[1]
  \State \LeftComment{0}{\Call{argsort}{$array$}: Returns the indices that would sort the given array.}
  \State \LeftComment{0}{\Call{getPermutation}{$signals$}: Returns the permutation number for the given array of signals.}
  \Procedure{PAM3\_SORT\_Encoding}{$\pamEightData$}
    \State $cnt(-1), cnt(0), cnt(1)\gets \Call{count}{\pamEightData}$
    \State \LeftComment{0}{$signals$ is an array of signals sorted according to the frequency.}
    \State $signals \gets \Call{argsort}{[cnt(-1), cnt(0), cnt(1)]}$ \label{alg_line:argsort}
    \State \LeftComment{0}{$permutation$ is a 3-bit permutation number.}
    \State $permutation \gets \Call{getPermutation}{signals}$ \label{alg_line:get_permu}
    \State $\pamEightData \gets \Call{Sort\_Convert}{\pamEightData, permutation$} \label{alg_line:sort_convert}
    \State\Return $\pamEightData, permutation$
  \EndProcedure
  \end{algorithmic}
\end{algorithm}

The PAM3-SORT encoding algorithm described in Algorithm \ref{alg:PAM3-SORT} is the algorithm that can achieve \emph{the least power consumption} among the three encoding algorithms we propose.

\figurename~\ref{fig:sort_convert} shows an example process of the he PAM3-SORT encoding algorithm. 
For each of two PAM-3 signal lines, we sort signals in ascending order through \textsc{argsort} at line \ref{alg_line:argsort} of Algorithm~\ref{alg:PAM3-SORT} based on the count values, $cnt(-1)$, $cnt(0)$, and $cnt(+1)$.
Next, we get the permutation number for the sorted signals through \textsc{getPermutation} at line \ref{alg_line:get_permu} of Algorithm~\ref{alg:PAM3-SORT}. There are three \textsc{PAM-3 signal values}, -1, 0, and +1 in total, and there are 3! (= 6) ways to permute the signals. After the conversion through the \textsc{PAM3\_SORT\_Encoding} algorithm proceeds, a 3-bit number representing the permutation mapping is returned.
In this example, the -1 signal is the least frequent, +1 is the next, and 0 is the most frequent before the conversion; thus, the permutation [-1, +1, 0] (assigned a permutation number 3).
During conversion, -1, +1, and 0 are converted into -1, 0, and 1, respectively.

\subsection{Discussion on Data Representation in PAM3-SORT}
\label{sec:data_representation}

To represent the binary data in PAM-3 signals, we convert three 8-bit words (24 bits in total) into two PAM-3 signal lines (8 signals per line).
Specifically, each binary symbol from three words (3 bits in total) can be efficiently represented by two consecutive PAM-3 signals of $cnt(-1)$, $cnt(0)$, or $cnt(+1)$.
This is one of the major differences that distinguish this work from~\cite{su2022energy}. 
In addition, our new PAM3-SORT requires only 5 bits ($2^5 = 32$) to represent 24 cases while the PAM4-SORT encoding method implemented in~\cite{su2022energy} used a table representing 4!, a total of 24 cases, to sort $cnt(00)$, $cnt(01)$, $cnt(10)$, and $cnt(11)$ in order of frequency.

In PAM-3, there are eight possible binary symbols in total $000$, $001$, $010$, $011$, $100$, $101$, $110$ and $111$ as shown in \figurename~\ref{fig:pam3_signal_modulation}.
If we naively apply the same representation for PAM4-SORT of the previous work~\cite{su2022energy} to PAM3-SORT, we will need a table with 40,320 ($=$8!) entries for all possible permutations of signal mapping, which requires 16 bits ($2^{16}=65,536$) to index all entries. That is, when we use PAM3-SORT encoding, additional 16 bits must be transmitted together with the data lines.
This overhead will enormously increase power consumption as well as the data processing time.

To address this problem, we represent these eight binary symbols in terms of three PAM-3 signals, $cnt(-1)$, $cnt(0)$, and $cnt(+1)$, after converting the binary symbols into PAM-3 signals. Then, the mapping table required to run PAM3-SORT will need only 6 ($=3!$) entries, and the indices of entries can be represented with only three bits ($2^3 = 8$).
This 3-bit representation is considerably simpler than the naive 16-bit representation of eight symbols and also smaller than the 5-bit representation used by the PAM4-SORT encoding algorithm proposed by the previous work~\cite{su2022energy}.
Thus, our representation of PAM-3 signals can significantly reduce the number of bits needed as well as the processing time required for the conversion process using the enormous mapping table.

\section{Evaluation}


\begin{table}
\centering
\caption{\label{tab:hw-table}\textsc{Hardware Implementation Costs For Each Encoding Algorithm}}
\normalsize
\scalebox{0.7}{
\begin{tabular}{|c|ccc|}
\hline \textbf{\begin{tabular}[c]{@{}c@{}}Encoding Algorithm\end{tabular}} & \multicolumn{1}{c|}{\textbf{PAM3-DBI}} & \multicolumn{1}{c|}{\textbf{PAM3-MF}} & \textbf{PAM3-SORT}
\\ 
\hline \textbf{\begin{tabular}[c]{@{}c@{}}Data width (bits)\end{tabular}} & \multicolumn{1}{c|}{24x2} & \multicolumn{1}{c|}{24x2} & 24x2
\\
\hline \textbf{\begin{tabular}[c]{@{}c@{}}Flag wires (bits)\end{tabular}} & \multicolumn{1}{c|}{1} & \multicolumn{1}{c|}{2} & 3
\\
\hline \textbf{\begin{tabular}[c]{@{}c@{}}Required \\ hardware logic\end{tabular}} & \multicolumn{1}{c|}{\begin{tabular}[c]{@{}c@{}}3 Counters, \\ Signal \\ inverters\end{tabular}} & \multicolumn{1}{c|}{\begin{tabular}[c]{@{}c@{}}3 Counters, \\ Signal \\ swapper\end{tabular}} & \begin{tabular}[c]{@{}c@{}}3 Counters, \\ Signal sorter, \\ Signal swapper\end{tabular}
\\ \hline 
\end{tabular}}
\label{tab:hw_table}
\end{table}

\begin{table}
\centering
\caption{Experimental Architecture Configurations}
\normalsize
\scalebox{0.7}{
\begin{tabular}{|l|l|}
\hline \textbf{Processor} & Out of order (O3)                                          \\ 
\hline \textbf{Cache} & \begin{tabular}[c]{@{}l@{}}L1: I-cache(32kB)  D-cache(64kB) L2(2MB)\end{tabular} 
\\ 
\hline \textbf{DRAM} & DDR4\_2400\_16x4, 1 channel, 512MB                              \\
\hline
\end{tabular}}
\label{tab:archi_config}
\end{table}
\subsection{Hardware Implementation Costs}
\label{sec:hw-costs}

\tablename~\ref{tab:hw_table} shows the hardware implementation costs we estimate for PAM3-DBI, PAM3-MF, and PAM3-SORT. 

Each algorithm receives two lines of PAM-3 signals, each containing eight signals generated from 24 bits.
PAM3-DBI needs a 1-bit flag to indicate the signal inversion, PAM3-MF needs a 2-bit flag for the most frequent signal, and PAM3-Sort needs a 3-bit flag for the permutation number.
All require 3 hardware counters to count each PAM-3 signal in the data word.
For signal conversion, each algorithm needs different hardware.
PAM3-DBI requires signal inverters.
PAM3-MF and PAM3-SORT algorithms require a signal swapper.
The PAM3-SORT algorithm also requires a signal sorter logic to sort three PAM-3 signals in the order of signal frequency.

\subsection{Experimental Setup}
\label{sec:exp-set}

We conducted experiments using similar environments as those used in Su \textit{et al.}~\cite{su2022energy}, with the same processor, cache, and DRAM configured by the gem5~\cite{binkert2011gem5} simulator as shown in \tablename~\ref{tab:archi_config}. 
We used DRAM read/write access traces extracted from the gem5 architecture simulator executing MiBench benchmark programs~\cite{guthaus2001mibench} on ARM and x86 ISAs.






Then, we used DRAM read/write traces from gem5 simulations as input to our Python script that implements our PAM-3 encoding algorithms to compute power consumption.
To use the binary traces as PAM-3 signals, we converted three 8-bit binary words (24 bits in total) into two PAM-3 signal lines (8 signals/line) as discussed in Section~\ref{sec:data_representation}.
Using these PAM-3 signal lines, we compared the power consumption of our encoding algorithms relative to the power consumption of the baseline PAM-3 signaling without our proposed algorithms.

\begin{figure}
\centering
\includegraphics[width=0.8\columnwidth]{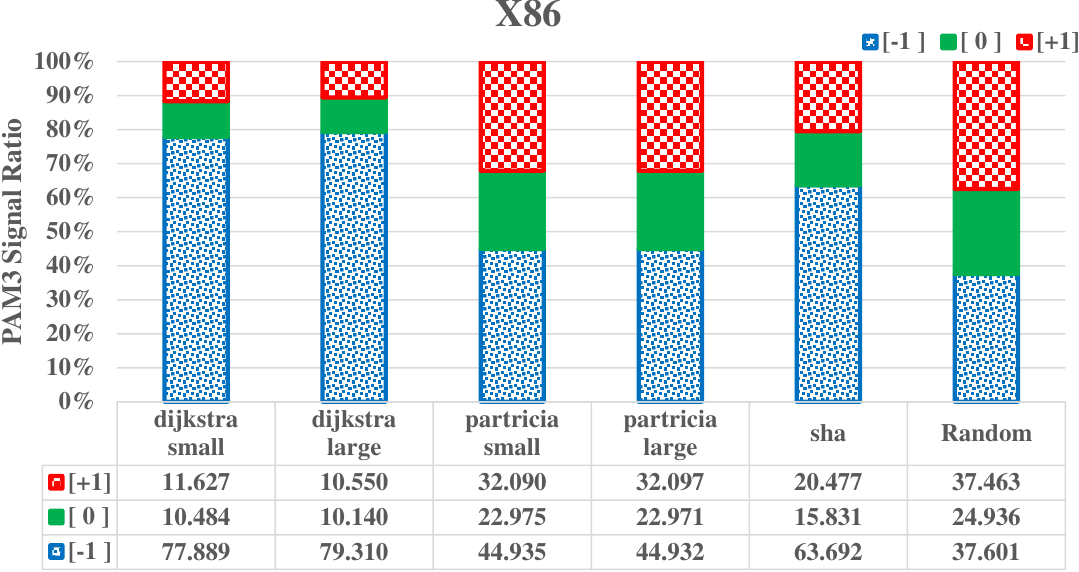}
\caption{The distribution of each PAM-3 signal (-1, 0, or 1) from the memory traces of MiBench programs executed on x86 ISA and from random bits.}
\label{fig:signal_percentage_X86}
\end{figure}

We also estimated the distribution of the PAM-3 signals for each benchmark program and randomly generated binary sequences.
The results on x86 are shown in \figurename~\ref{fig:signal_percentage_X86}.
We found that $-1$ signals are dominant, with similar results on ARM.


\subsection{Experimental Results}
\label{sec:exp-result}

For evaluation, we defined the Termination Power Consumption Ratio shown in the Y-axis of \figurename~\ref{fig:termination_power_percentage_ARM}, the percentage of termination power consumption of each PAM-3 encoding algorithm ($TerminationPower_{PAM3}$) compared to the case where the PAM-3 encoding algorithm was not applied ($TerminationPower_{Original}$) in the following equation:

\begin{figure}
\centering
\includegraphics[width=0.91\columnwidth]{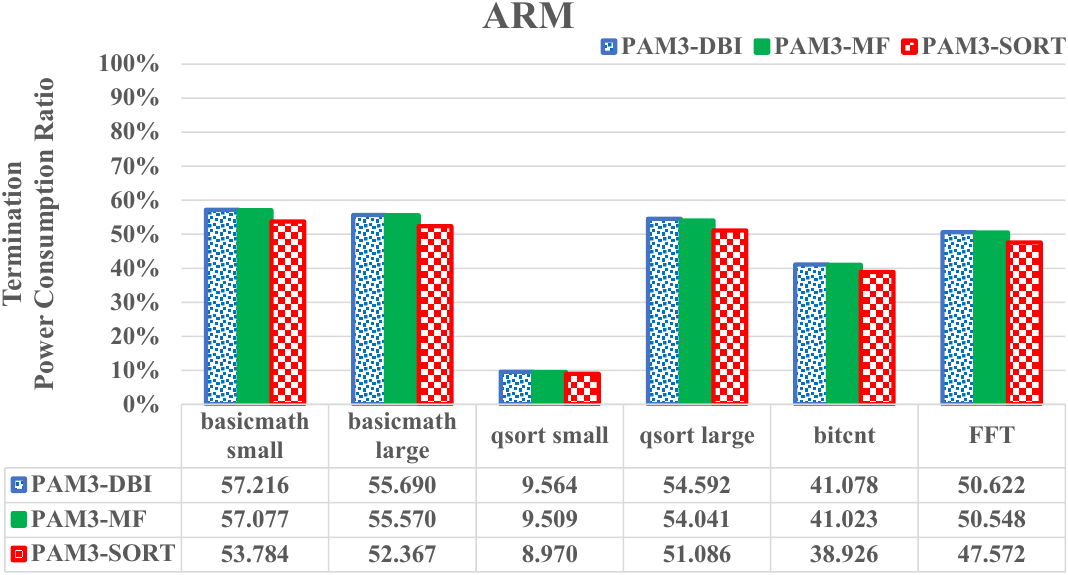}
\caption{Termination power consumption ratio, explained in equation (\ref{eq:termination_power_ratio}), for each encoding algorithm and MiBench programs executed on ARM ISA. (The lower ratio indicates more termination power reduction achieved.)}
\label{fig:termination_power_percentage_ARM}
\end{figure}

\begin{figure}
\centering
\includegraphics[width=0.91\columnwidth]{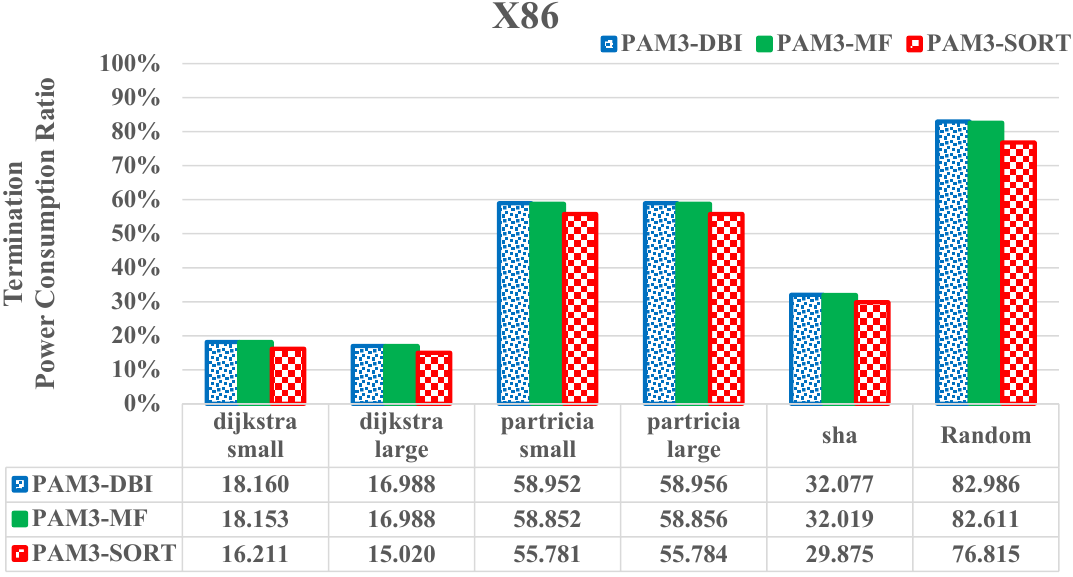}
\caption{Termination power consumption ratio for each encoding algorithm and MiBench programs executed on x86 ISA.
}
\label{fig:termination power percentage_X86}
\end{figure}

\begin{figure}
\centering
\includegraphics[width=0.91\columnwidth]{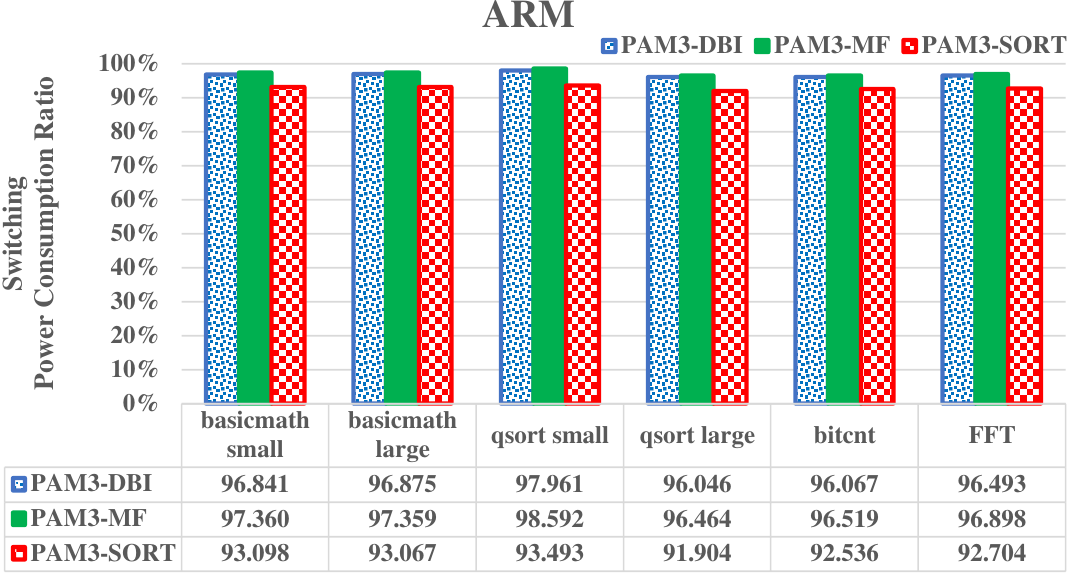}
\caption{Switching power consumption ratio for each encoding algorithm and MiBench programs executed on ARM ISA. (The lower ratio indicates more switching power reduction achieved.)}
\label{fig:switching power percentage_ARM}
\end{figure}

\begin{figure}
\centering
\includegraphics[width=0.91\columnwidth]{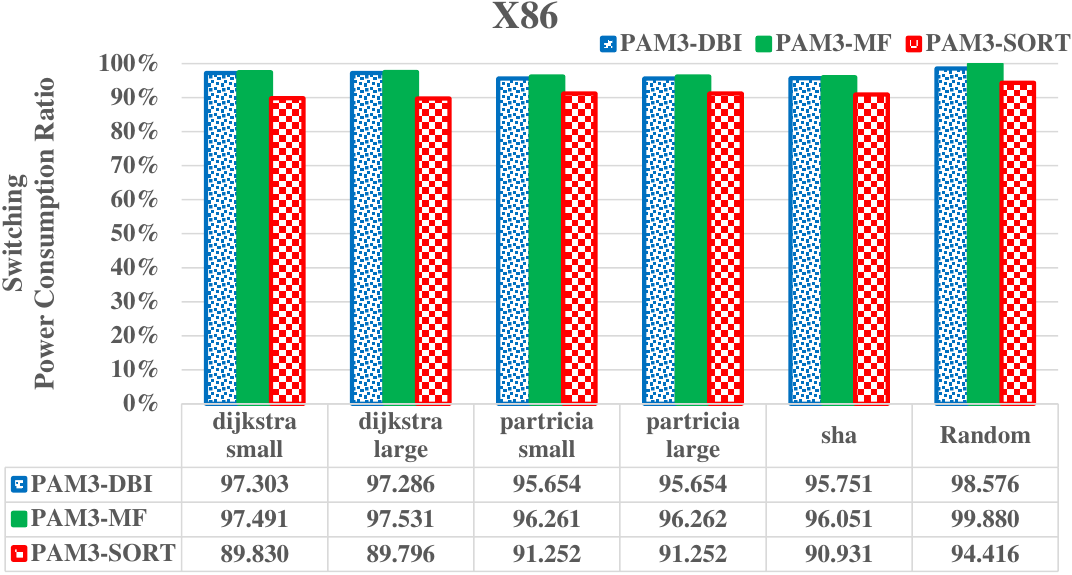}
\caption{Switching power consumption ratio for each encoding algorithm and MiBench programs executed on x86 ISA. 
}
\label{fig:switching power percentage_X86}
\end{figure}

\begin{equation}
\footnotesize
Ratio_{TerminationPower}(\%) = \frac{TerminationPower_{PAM3}}{TerminationPower_{Original}}\times100(\%)
\label{eq:termination_power_ratio}
\end{equation}



Next, we compared the termination power consumption of the three algorithms against the base as shown in \figurename~\ref{fig:termination_power_percentage_ARM} and \figurename~\ref{fig:termination power percentage_X86}.
All three proposed algorithms achieved a significant reduction in termination power, ranging from 41\% reduction with PAM3-DBI running patricia large on x86 to 90\% reduction with PAM3-SORT running qsort small on ARM.

Our results show that the termination power consumption of the PAM3-DBI and PAM3-MF algorithms is almost the same, while PAM3-SORT has the lowest termination power consumption among the three proposed encoding algorithms.
This is because both PAM3-DBI and PAM3-MF algorithms replace the -1 signals with the +1 signals at a similar rate, as PAM3-DBI kicks in when there are more -1 signals than +1 signals, while PAM3-MF replaces the most frequent signal, which is often -1, with the +1 signals, while PAM3-SORT remaps all three signals in the order of signal frequency.  


In addition to the termination power, we also evaluated the switching power of the proposed algorithms.
 The switching power consumption ratio for each algorithm is shown in \figurename~\ref{fig:switching power percentage_ARM} and \figurename~\ref{fig:switching power percentage_X86}.
The evaluation results clearly show that all three algorithms consume less switching power compared to the baseline without the proposed algorithms.


\section{Conclusion}
In this paper, we propose our novel energy-efficient encoding methods for the PAM-3 DRAM bus, \emph{PAM3-DBI}, \emph{PAM3-MF}, and \emph{PAM3-SORT}.
The evaluation results with benchmark programs clearly show that our proposed encoding algorithms achieve a significant reduction in termination power compared to the baseline.
Our proposed algorithms also showed less switching power consumption compared to the baseline.
Our analysis of hardware usage of the proposed algorithms suggests that \emph{PAM3-DBI} will cost the least additional hardware.

\section*{Acknowledgments}
This work was supported in part by Institute of Information \& communications Technology Planning \& Evaluation (IITP) grants funded by the Korea government (MSIT) (IITP-2024-RS-2023-00253914 and No.2021-0-00754).

\bibliographystyle{IEEEtran}
\bibliography{refs.bib}

\vfill

\end{document}